\documentclass{article}
\usepackage[dvips]{graphicx}
\usepackage{cite}
\def\1ad{\mbox{\normalsize $^1$}}
\def\2ad{\mbox{\normalsize $^2$}}
\def\3ad{\mbox{\normalsize $^3$}}
\def\4ad{\mbox{\normalsize $^4$}}
\def\5ad{\mbox{\normalsize $^5$}}
\def\6ad{\mbox{\normalsize $^6$}}
\def\7ad{\mbox{\normalsize $^7$}}
\def\8ad{\mbox{\normalsize $^8$}}
\def\makefront{\vspace*{1cm}\begin{center}
\def\newtitleline{\\ \vskip 5pt}
{\Large\bf\titleline}\\
\vskip 1truecm
{\large\bf\authors}\\
\vskip 5truemm
\addresses
\end{center}
\vskip 1truecm
{\bf Abstract:}
\abstracttext
\vskip 1truecm}
\begin {document}
%
%
\def\titleline{%
Center dominance, Casimir scaling, and confinement
\newtitleline
in lattice gauge theory
\protect\footnote{Based on talks presented by
\v{S}.\ Olejn\'{\i}k 
at the 31st International
Symposium on the Theory of Elementary Particles,
Buckow, Germany, September 2--6, 1997, and at the 7th Workshop on Lattice 
Field Theory, Vienna, Austria, September 18--20, 1997.
Work supported in part by the Slovak GAS, Grant No.\ 2/4111/97.}
}
%
%
\def\authors{%
L.\ Del Debbio\ \1ad, M.\ Faber\ \2ad, J.\ Greensite\ \3ad, 
\v{S}.\ Olejn\'{\i}k\ \4ad
}
%
%
\def\addresses{%
\1ad
Dept.\ of Physics and Astronomy, University of
Southampton, Southampton SO17 1BJ, UK\\
\2ad
Institut f{\"u}r Kernphysik, Technische Universit\"at 
Wien, A--1040 Vienna, Austria\\
\3ad
The Niels Bohr Institute, DK--2100 Copenhagen \O, Denmark\\
\4ad
Institute of Physics, Slovak Academy of Sciences, SK--842 28 
Bratislava, Slovakia
}
%
%
\def\abstracttext{%
We present numerical evidence that supports the theory
of quark confinement based on center vortex condensation.
We introduce a special gauge (``maximal center gauge'')
and center projection,
suitable for identification of center vortices.
Main focus is then put  on the connection of vortices in
center projection to ``confiners'' in full, unprojected
gauge-field configurations. Topics briefly discussed include: the
relation between vortices and monopoles, first results for SU(3), and
the problem of Casimir scaling. 
}
%
%
%
%
\makefront
%
%
%
%
\section{Introduction}
%
The most popular model of colour confinement in QCD
relies on the idea of ``dual superconductivity'', due to `t~Hooft and 
Mandelstam. A realization of the idea is the abelian-projection
theory of `t~Hooft~\cite{abelproj}: 
he suggested to fix to an ``abelian projection'' gauge,
reducing the SU($N$) gauge symmetry to U(1)$^{N-1}$, and 
identifying abelian gauge
fields (with respect to the residual symmetry) and magnetic monopoles.
Abelian electric charges then become confined due to monopole condensation.
In 1987, Kronfeld et al.~\cite{Kronfeld}
suggested testing `t~Hooft's theory in
lattice simulations, in a special gauge that makes SU($N$) link variables
as diagonal as possible. If one computes various physical observables
using the diagonal parts of the links only, one observes ``abelian''
dominance~\cite{dominance}: 
the expectation values of the physical quantities in the full
non-abelian theory (often) coincide with the ones in the abelian theory 
obtained by the abelian projection in the maximal abelian gauge.
Much evidence has been obtained for the abelian-projection picture
and the model of dual superconductivity, 
but there remain problems to be solved. We have underlined its inability
to explain approximate Casimir scaling of the linear potential between
higher-representation colour sources at intermediate length 
scales~\cite{PRD96,lat96}.

Another picture of confinement was quite popular before the advent of
dual superconductivity, namely the $Z_N$ vortex condensation theory,
proposed, in various forms, by many authors~\cite{vortex}. 
According to this model, the QCD vacuum is 
filled with vortices, having the topology of tubes (in 3 Euclidean dimensions)
or surfaces (in 4 dimensions) of finite thickness, which carry magnetic flux
quantized in terms of elements of the center of the gauge group.
Center vortices are assumed to condense in the QCD vacuum. The area-law
fall-off of large Wilson loops comes from fluctuations in the number
of center vortices linked to the loops.

The $Z_N$ vortex condensation theory apparently suffers from the same
``Casimir-scaling disease'' as the abelian projection does: there seems
no way of accommodating the existence of a linear potential between 
adjoint sources to the idea of vortices dominating the QCD vacuum.
There exists a simple solution to this controversy~\cite{castex}, and we
will discuss it at the end of this paper.

With perhaps only one exception~\cite{Tomboulis}, 
the ideas behind the 
vortex-condensation picture have not in the past been subjected to lattice 
tests. The aim of our investigation is to study the vortex theory in numerical
Monte Carlo simulations, by methods and approaches inspired to some
extent by earlier work of many authors in the abelian projection theory.

\section{Maximal center gauge, projection and dominance}
%
The maximal abelian gauge underscores the role of the largest abelian
subgroup of the gauge group. In much the same way, one can choose a gauge
condition in which the gauge group {\em center\/} is given prominent 
importance.
In SU(2) lattice gauge theory we proposed~\cite{lat96,PRD97} 
to fix to {\em maximal center
gauge\/} (MCG) by making link variables $U$ as close as possible to its center
elements $\pm I$. There are many (in fact infinitely many) ways how to do
it; we implemented two simple choices:

1.\ The {\em indirect maximal center gauge\/} 
(IMCG)~\cite{lat96,PRD97,Zako,lat97}: We first fix to maximal
abelian gauge (MAG) in the usual way, by maximizing the quantity
\begin{equation} 
\sum_x\sum_\mu \mbox{\rm Tr}\;[U_\mu(x)\sigma_3 U_\mu^\dagger(x)\sigma_3],
\end{equation}
then extract from $U_\mu(x)$ their diagonal parts 
$A_\mu(x)=\exp[i\theta_\mu(x)\sigma_3]$,
and use the remnant U(1) symmetry to bring $A_\mu(x)$ as close as possible to
the center elements by maximizing
\begin{equation}
\sum_x\sum_\mu \;\cos^2\theta_\mu(x).
\end{equation}

2.\ The {\em direct maximal center gauge\/} (DMCG)~\cite{DMCG,Zako}: 
To fix this gauge
one looks directly for the maximum of
\begin{equation}
\sum_x\sum_\mu \mbox{\rm Tr}\;[U_\mu(x)] \mbox{\rm Tr}\;[U_\mu^\dagger(x)].
\end{equation}
%

%
%
\begin{figure}
\begin{center}
\begin{tabular}{c c c}
{\scalebox{.28}[0.31]{\rotatebox{90}{\includegraphics{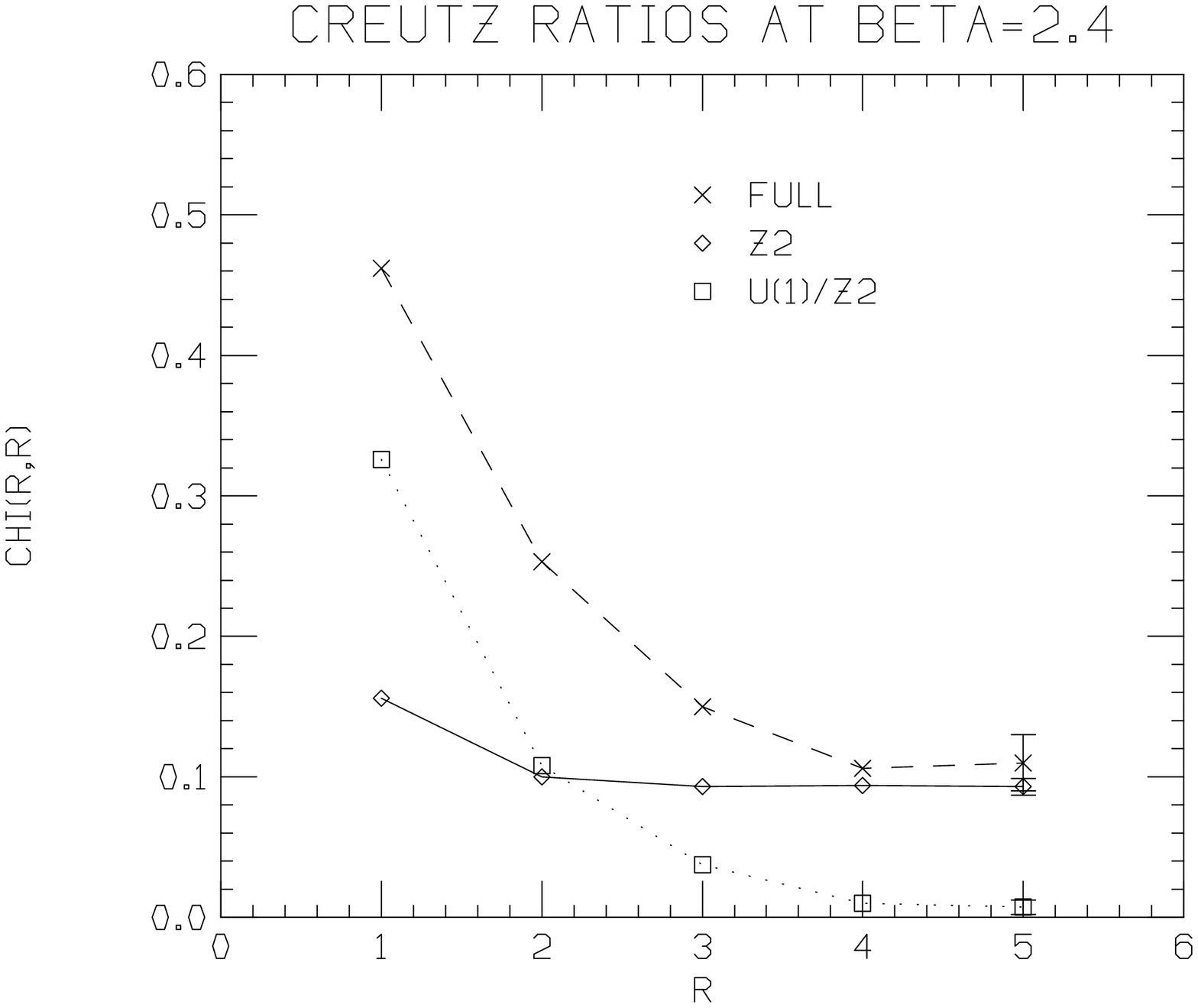}}}}&~~~~~~&
{\scalebox{.40}[0.3]{\rotatebox{90}{\includegraphics{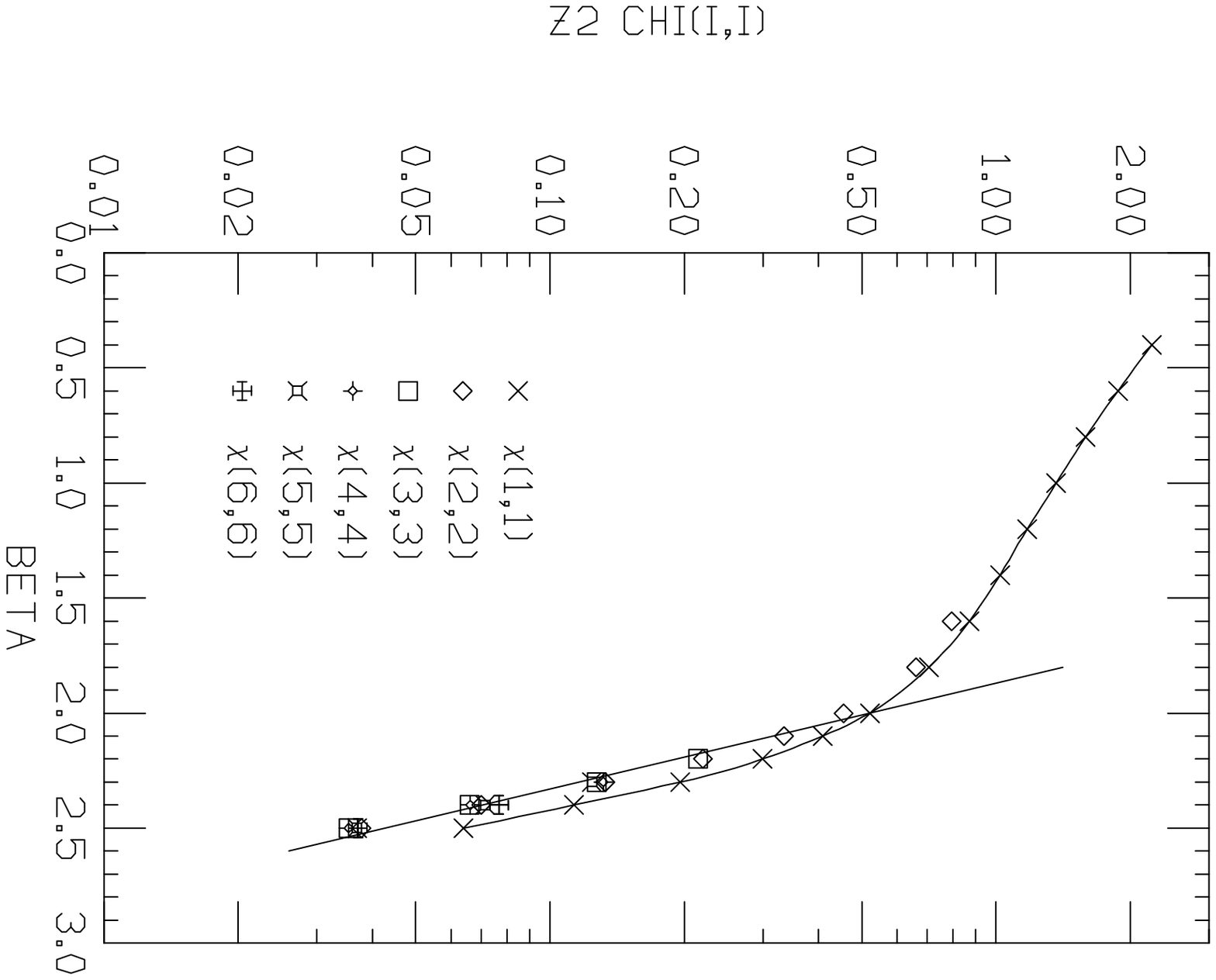}}}}\\
{\tiny\hspace*{1cm}(a)}&~~~~~~&{\tiny\hspace*{1cm}(b)}
\end{tabular}
\caption{Creutz ratios: (a) vs.\ $R$ for full, center-projected, and 
U(1)/$Z_2$-projected lattice configurations at $\beta=2.4$ (IMCG),
(b) vs.\ $\beta$ from center-projected configurations (DMCG). The straight 
line is the asymptotic freedom prediction: $\sigma a^2 = (\sigma/\Lambda^2)
\left(6\pi^2\beta/11\right)^{102/121}\exp\left[-6\pi^2\beta/11\right]$.}
\label{fig1}
\end{center}
\end{figure}

In both cases, after gauge fixing one is left with the remnant $Z_2$
gauge symmetry.\footnote{Qualitatively, the same physical results are 
obtained in both gauges (cf.~\cite{PRD97,DMCG}).}

The next step is {\em center projection\/}, i.e.\ replacing
full link matrices $U$ (in a particular MCG) by  center elements $Z$,
which are defined to be
\begin{equation}
Z\equiv \mbox{\rm sign}\;(\cos(\theta))\;I \quad\mbox{\ (in IMCG)}
\qquad\mbox{or}\qquad
Z\equiv \mbox{\rm sign}\;(\mbox{\rm Tr}\;(U))\;I \quad\mbox{\ (in DMCG),}
\end{equation}
and to compute various physical quantities of interest, e.g.\ Wilson loops
and Creutz ratios, using the $Z$ links.

Figure~\ref{fig1}a compares Creutz ratios $\chi(R,R)$ at $\beta=2.4$
computed from full lattice configurations and from center-projected
configurations. We clearly see {\em center dominance\/}: Creutz ratios
computed from $Z$ links agree with full Creutz ratios at large enough
distances, the asymptotic values of the string tension almost coincide.
On the contrary, Creutz ratios computed from links with the $Z$ variable 
factored out (dotted line in Figure~\ref{fig1}a) show no string tension 
at all. Another interesting observation is that the Creutz ratios
computed from center-projected configurations almost do not depend on $R$;
center projection removes Coulombic contributions.

In Figure~\ref{fig1}b we plot Creutz ratios vs.\ $\beta$, extracted from 
center-projected configurations in DMCG. The straight line is the
asymptotic freedom prediction with the value of $\sqrt{\sigma}/\lambda=58$,
which very well agrees with ``state-of-the-art'' asymptotic string tension
computations~\cite{Bali}.

Our data show that the $Z$ center variables are crucial parts of the $U$
links in MCG, in particular they carry most of the information
on the string tension. This phenomenon of {\em center dominance\/} gives
rise to a whole series of questions on the role and nature of center vortex
configurations in the QCD vacuum. We  will list a few of those questions here
and sketch our tentative answers.

\section{Questions and answers}
%
\subsection{Vortices and confinement?}
%
\begin{description}
\item
{\em Question 1:\/}
Has center dominance anything to do with confinement? What, if any,
is the relation of $Z_2$ vortices seen after center projection to
``confiners'' in full, unprojected configurations?
\end{description}

%
%
\begin{figure}[t!]
\begin{center}
\begin{tabular}{c c}
{\scalebox{.40}{{\includegraphics{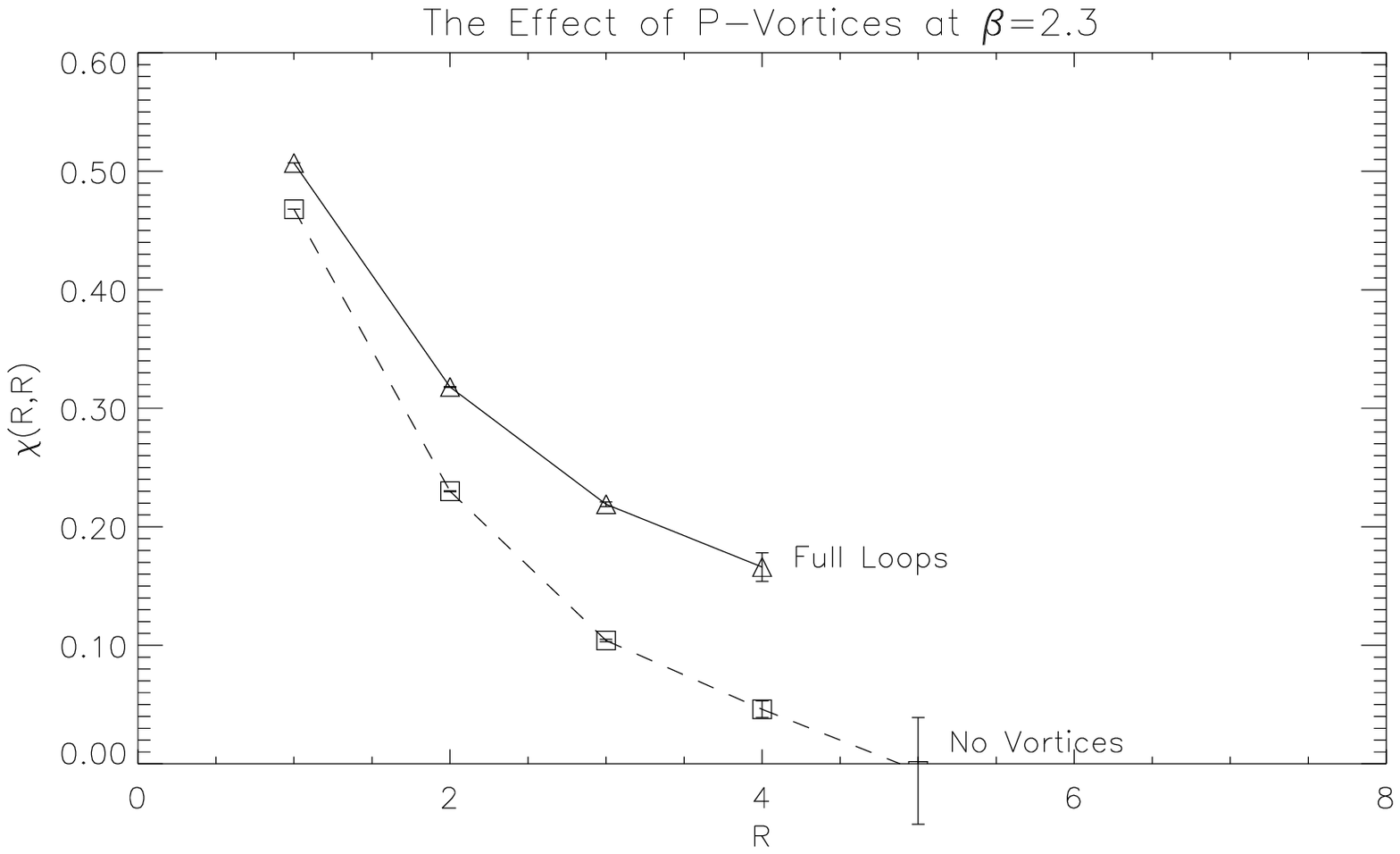}}}}&
{\scalebox{.40}{{\includegraphics{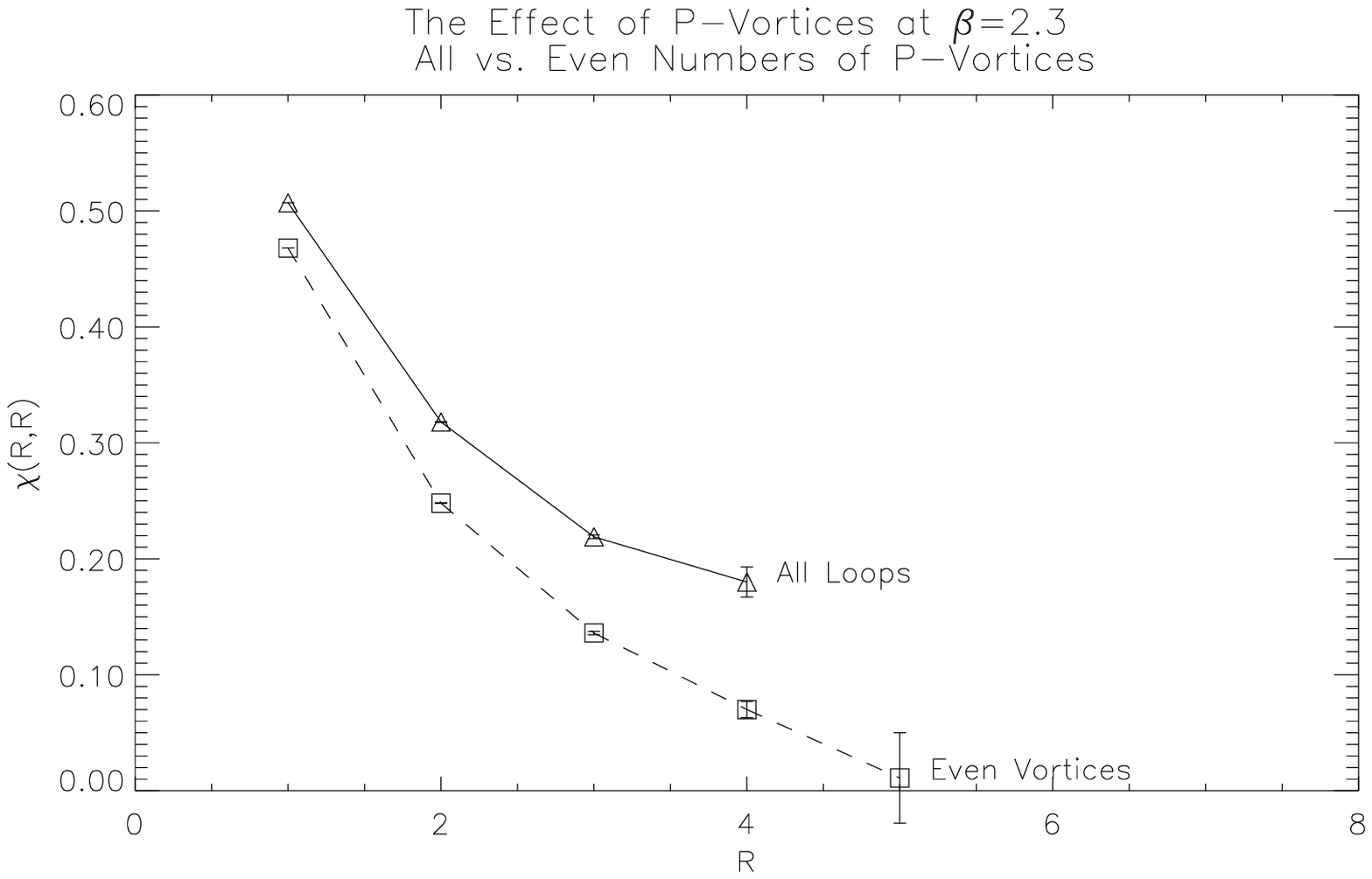}}}}\\
{\tiny\hspace*{1cm}(a)}&{\tiny\hspace*{1cm}(b)}
\end{tabular}
\caption{Creutz ratios extracted from loops with (a) no P-vortices,
(b) even number of P-vortices piercing the loop, compared with the usual
Creutz ratios at $\beta=2.3$.}
\label{fig2}
\end{center}
\end{figure}

To answer the question, we first introduce the notion of a {\em
P(rojection)-vortex\/}. The excitations of a $Z_2$ lattice gauge theory with 
non-zero action are ``thin'' vortices, having the topology of a surface,
one lattice spacing thick. We will call such vortices in center projected
$Z$-link configurations P-vortices. A plaquette is pierced by a P-vortex if,
after maximal center gauge fixing and center projection, the corresponding
projected plaquette has the value of $-1$.

However, we want to emphasize that center projection, and abelian
projection as well, represents an uncontrollable truncation
of full  lattice configurations. Therefore we will not base our following
arguments on measurements in center projected
configurations. Instead,
we will use center projection mainly for selecting sub-ensembles of 
configurations on which physical quantities are evaluated.
In particular, we will compute $W_n(C)$, Wilson loops evaluated on such a 
sub-ensemble of configurations that precisely $n$ P-vortices, in the 
corresponding center-projected configurations, pierce the minimal area of the
loop. Though the data set is selected in center projection,
the Wilson loops
themselves are evaluated using the full, unprojected link variables.

From the computed vortex-limited Wilson loops $W_n(C)$ one can determine
Creutz ratios $\chi_n$. A simple test of a relation of P-vortices to
confinement is then the following: if the presence/absence of P-vortices
is {\em irrelevant\/} for confinement, then we would expect
$\chi_0(I,J)\approx\chi(I,J)$ for large loops.
The result of the test is shown in Figure~\ref{fig2}a. The string tension
vanishes if P-vortices are excluded from Wilson loops; it also vanishes if
only odd numbers of P-vortices are excluded (Fig.~\ref{fig2}b). 
The presence/absence  of P-vortices seems strongly correlated with 
the presence/absence of ``confiners'' in unprojected field configurations.

However, the true ``confiners'' do not necessarily have to be any sort
of $Z_2$ vortices. The natural question is, whether the objects identified
using center projection tend to carry $Z_2$ magnetic flux. A simple 
argument~\cite{PRD97,Zako} leads to the expectation that
${W_n(C)}/{W_0(C)}\rightarrow(-1)^n$ for large loops.

\begin{figure}[t]
\begin{center}
\begin{tabular}{c c}
{\scalebox{.40}{{\includegraphics{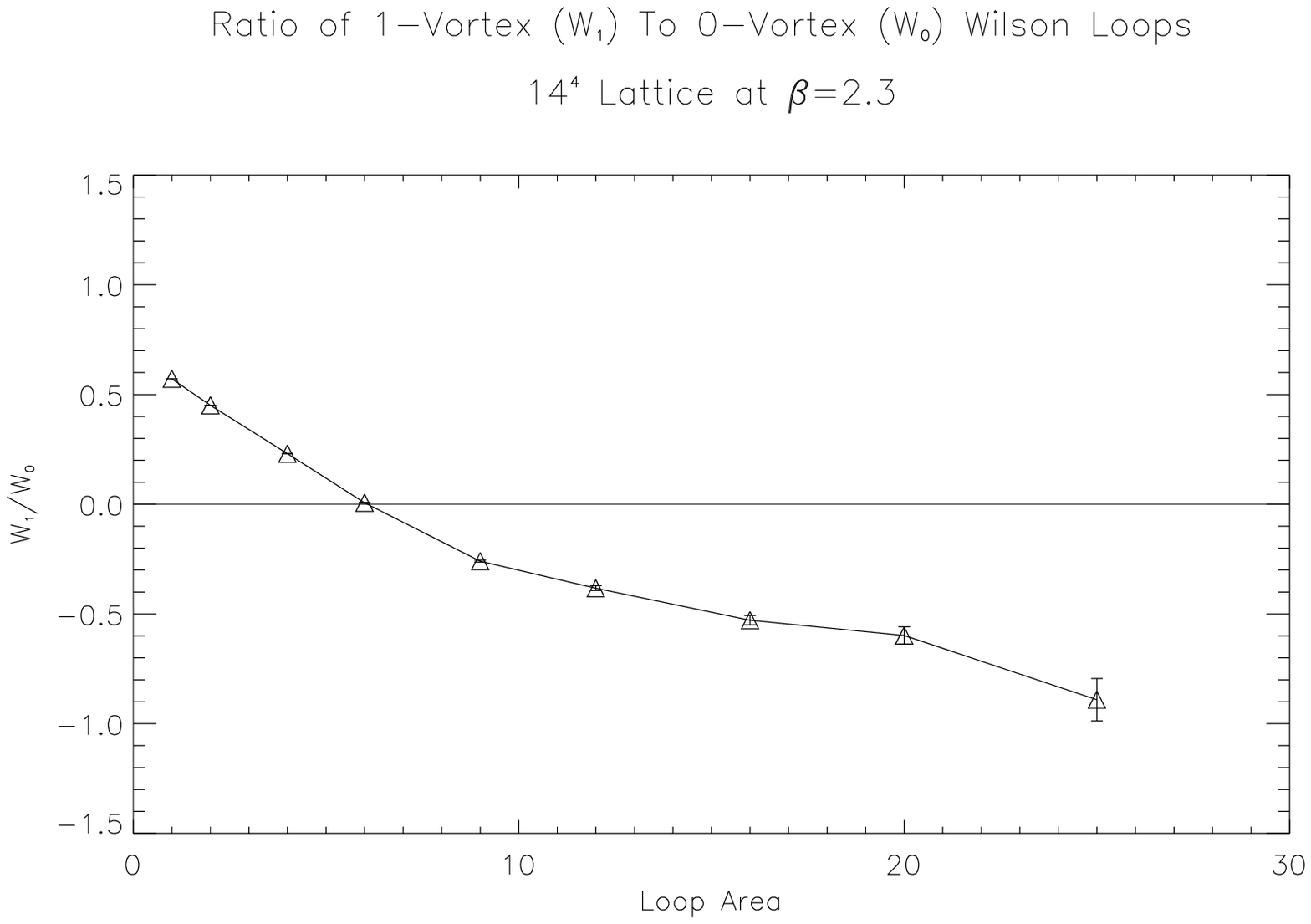}}}}&
{\scalebox{.40}{{\includegraphics{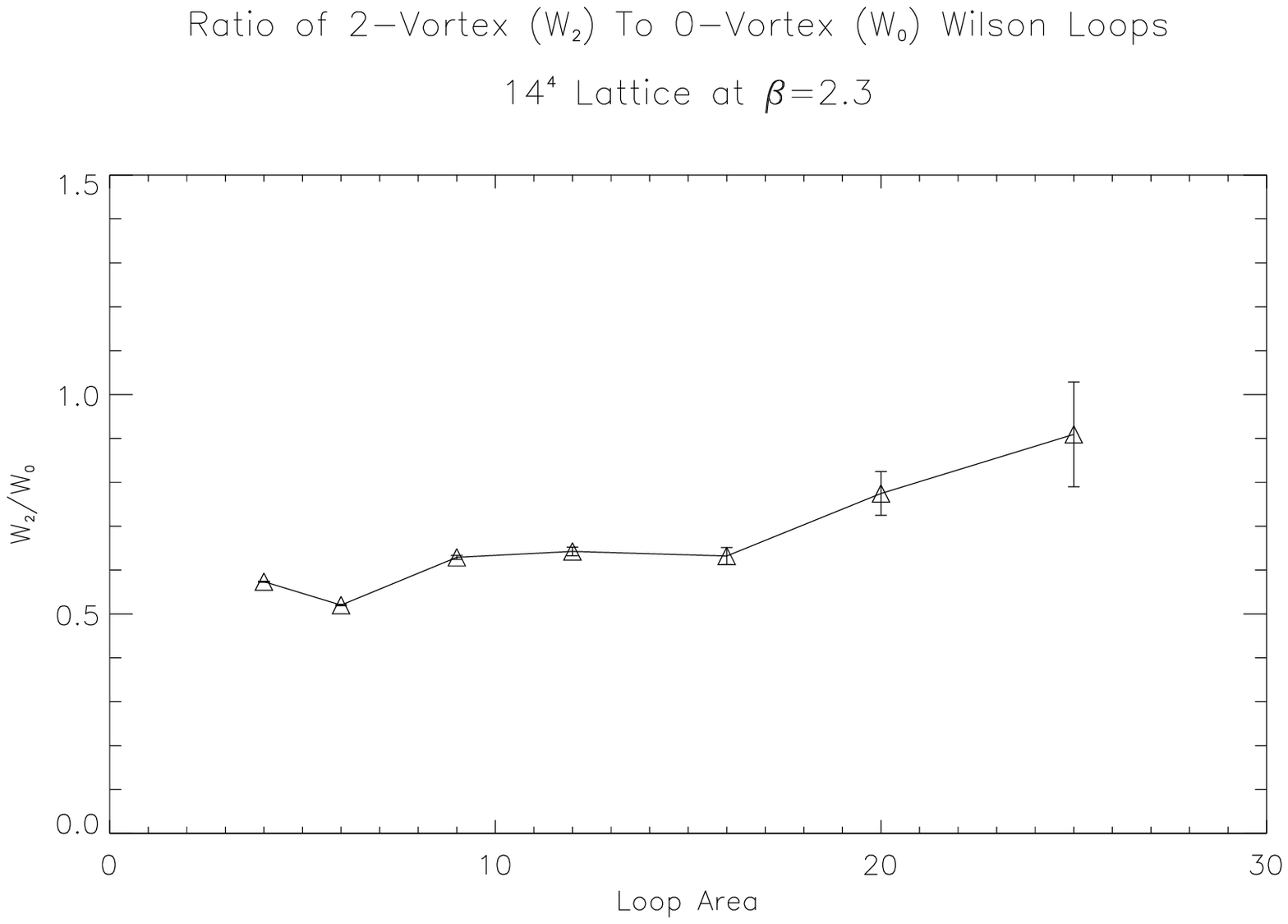}}}}\\
{\tiny\hspace*{1cm}(a)}&{\tiny\hspace*{1cm}(b)}
\end{tabular}
\caption{%
(a) $W_1(C)/W_0(C)$, (b) $W_2(C)/W_0(C)$ vs.\ loop area at $\beta=2.3$.}
\label{fig3}
\end{center}
\end{figure}

Figure~\ref{fig3} shows our data for $W_1/W_0$ and $W_2/W_0$
at $\beta=2.3$. They are consistent with the expectation and thus indicate
that the confining gauge field configurations are center vortices.
However, the values of $(-1)^n$ are reached for relatively {\em large\/}
loop areas. The objects corresponding to P-vortices then appear to be
rather ``thick'' $Z_2$ vortices. In Section \ref{Casimir} will this fact be
related to a simple explanation of approximate Casimir scaling. 

%
%

Now we are ready to formulate
\begin{description}
\item
{\em Answer 1:\/} 
We found evidence that center dominance in maximal
center gauge is a reflection of the presence of thick center vortices
in the unprojected configurations. Those vortices are identified as
thin P-vortices in center projection.
\end{description} 

\subsection{Vortices and/or monopoles?}
%
\begin{description}
\item
{\em Question 2:\/}
If the vacuum is dominated, at long wavelengths, by
$Z_2$ vortex configurations, then how do we explain the numerical
successes of abelian projection in maximal abelian gauge?
\end{description}

We believe that the question is answered in
the following way:
\begin{description}
\item
{\em (Probable) Answer 2:\/}
A center vortex configuration,  transformed to maximal abelian gauge and 
then abelian-projected, will appear as a
chain of monopoles alternating with antimonopoles.  
These monopoles essentially arise because of the projection; 
they are condensed because
the long vortices from which they emerge are condensed.
\end{description}

The support for the answer was given in much detail in~\cite{Zako}.
It is clear, however, that this question deserves further study.

\subsection{SU(3)?}
%
\begin{description}
\item
{\em Question 3:\/}
In nature quarks appear in three colours. Do the observed phenomena
survive transition from SU(2) to SU(3)?
\end{description}

The maximal center gauge in SU(3) gauge theory is defined as the gauge
which brings link variables $U$ as close as possible to elements of
its center $Z_3 = \lbrace e^{-2i\pi/3}I,\;I,\;e^{2i\pi/3}I\rbrace$. 
This can be achieved e.g.\ by maximizing the quantity
\begin{equation}\label{baryonlike}
\sum_x\sum_\mu\mbox{Re}\left(\left[\mbox{Tr}\;U_\mu(x)\right]^3\right).
\end{equation}

Fixing to the maximal center gauge in SU(3) gauge theory turns out to be 
much more difficult and CPU-time consuming than in the case of SU(2).
Therefore our simulations have until now been restricted to small lattice 
sizes and to strong coupling.

\begin{figure}
\begin{center}
\begin{tabular}{c c}
{\scalebox{.40}{\includegraphics{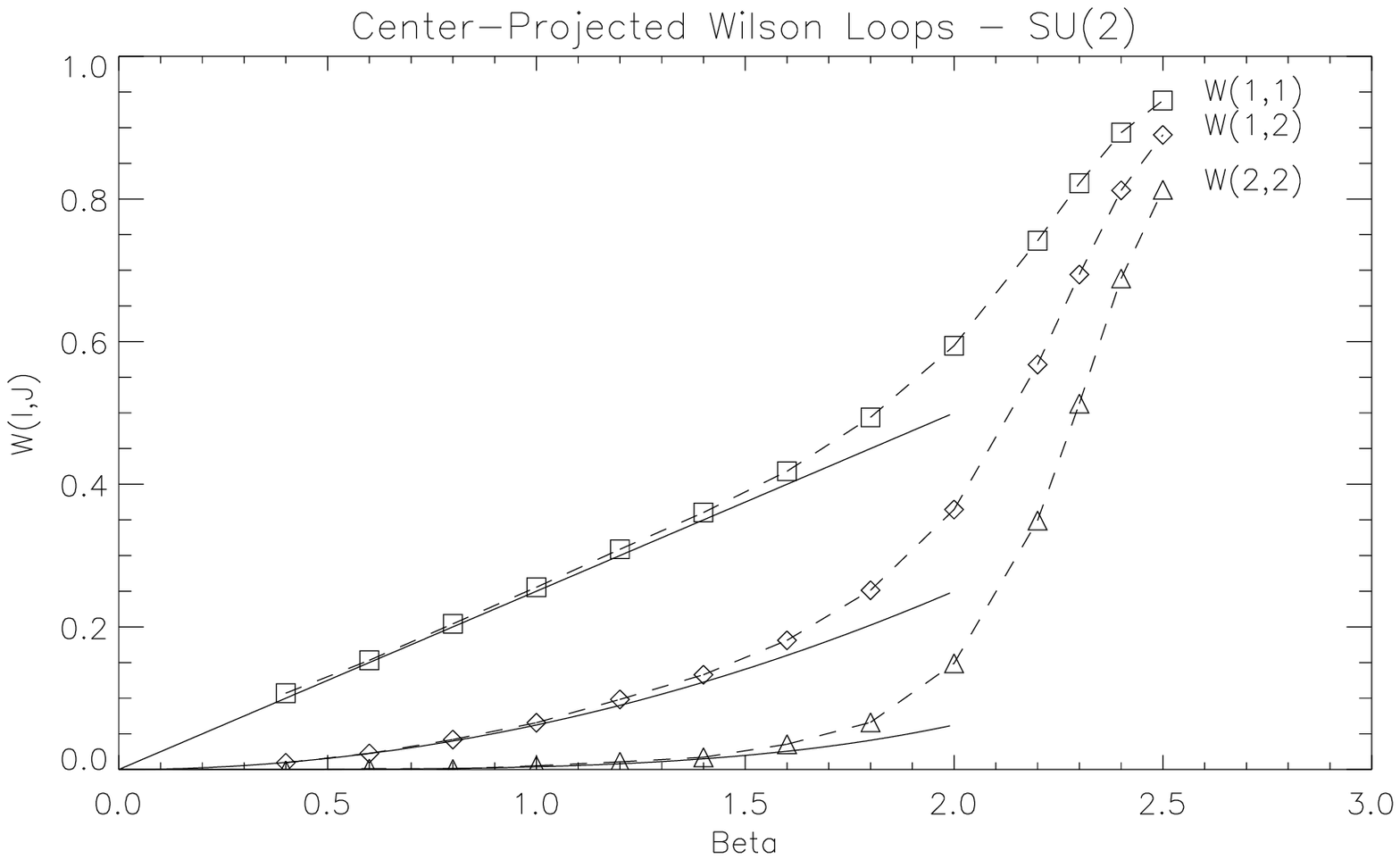}}}&
{\scalebox{.40}{\includegraphics{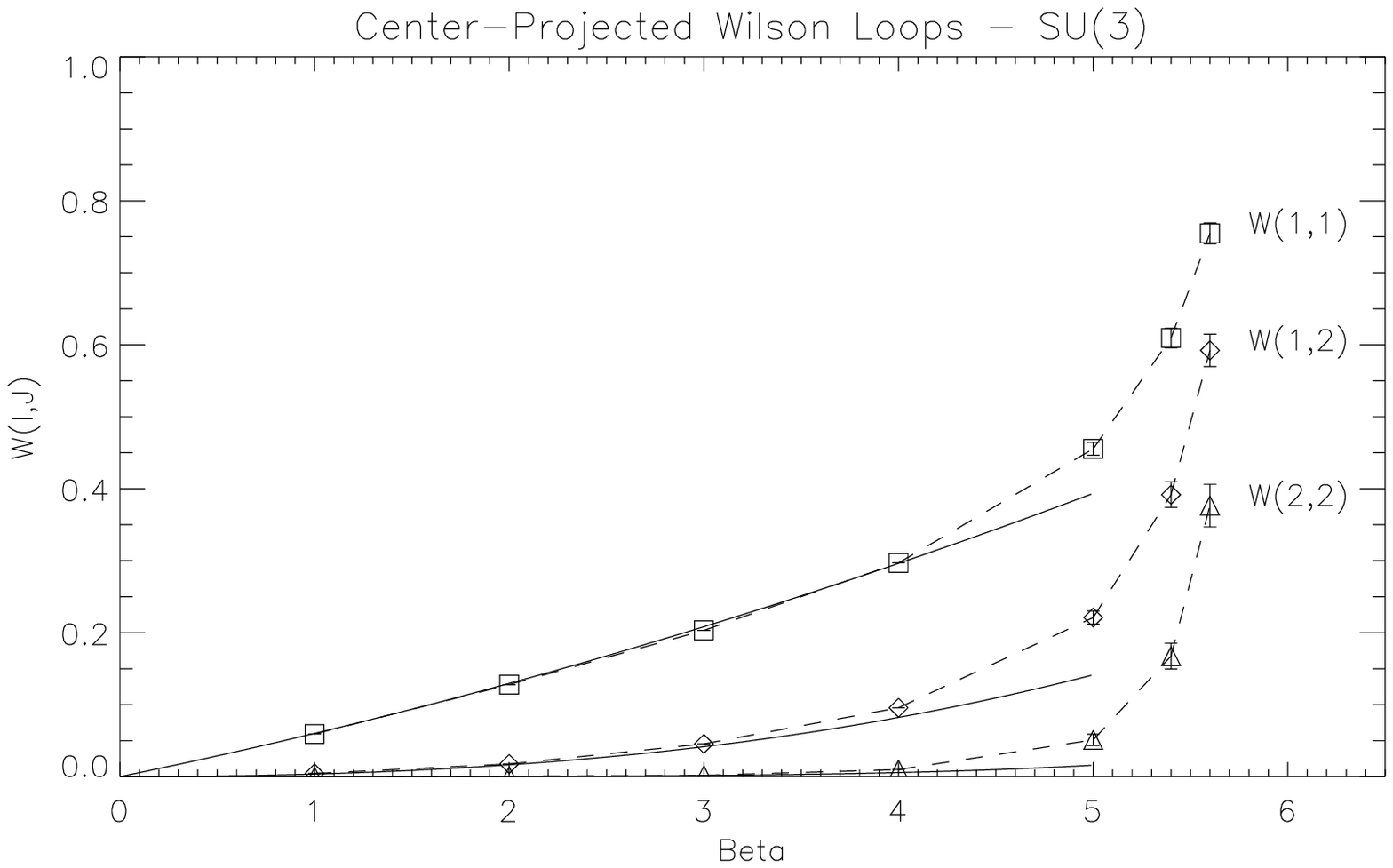}}}\\
{\tiny\hspace*{1cm}(a)}&{\tiny\hspace*{1cm}(b)}
\end{tabular}
\caption{Center-projected Wilson loops vs.\ the strong-coupling expansion
(solid lines) in SU(2) and SU(3) lattice gauge theory. SU(3) values
were obtained on an $8^4$ lattice.}
\label{fig4}
\end{center}
\end{figure}

Our strong coupling results for the SU(2) and SU(3) lattice gauge theory 
are compared in Figure~\ref{fig4}. 
In SU(2) Monte Carlo data agree with the strong coupling expansion up to
almost $\beta=1.5$.
Figure~\ref{fig4}b shows center-projected Wilson loops in SU(3)
together with results of strong-coupling expansion to leading and 
next-to-leading order.
The agreement extends up to $\beta=4$.

%
%

\begin{description}
\item
{\em (Partial) Answer 3:\/}
The situation at strong coupling looks much the same
in SU(2) and SU(3): in both cases full Wilson loops are well reproduced
by those constructed from center elements alone in
MCG. Thus, center dominance is seen in SU(3) gauge theory
at strong coupling.
\end{description}

\subsection{Casimir scaling?}\label{Casimir}
%
\begin{description}
\item
{\em Question 4:\/}
Is there any way of accommodating the approximate Casimir scaling
of higher-representation potentials to the vortex dominated 
QCD vacuum?
\end{description}

At the first sight, there is not. The adjoint representation
transforms trivially under the gauge group center, large adjoint Wilson loops
are unaffected by center vortices. As a result, the adjoint
string tension vanishes.

There is however a loophole in the above statements: Adjoint loops
are unaffected, unless the core of the vortex
happens to overlap with the perimeter of the loop. If, then, the
vortex thickness is quite large, on the order or exceeding the typical 
diameter of low-lying hadrons -- and our data seem to indicate the
presence of rather ``thick'' center vortices, -- the Wilson loops
can be influenced by vortices up to relatively large loop sizes.

A phenomenological model of the ``thick'' center vortex core has
been worked out in a recent paper of three of us~\cite{castex}. 
We cannot discuss the model in detail
here, we just mention that it leads to potentials between 
colour sources that show approximate
Casimir scaling at small and intermediate distances, and colour screening
of integer-representation sources at large distances. 

\begin{description}
\item
{\em Answer 4:\/}
The Casimir scaling of the string tensions of 
higher-representation Wilson loops is an effect due to the finite
(and large) thickness of the center vortex cores (see also \cite{Corn2}).%
\end{description} 

\section{Conclusions}
%
We subjected the picture of quark confinement based on center
vortices to simple tests on the lattice. We proposed a method of
localizing center vortices in thermalized lattice configurations,
and found vortices to be responsible for the asymptotic string tension
in SU(2) lattice gauge theory. The same holds also for SU(3) at strong
coupling.

Further, we posed the lattice a lot of simple questions on the nature
of vortex configurations. The tests indicate that the ``confiners''
in QCD are center vortices, and monopoles appear along vortices as artifacts
of abelian projection. It is tempting to believe that monopole 
condensation might be just a manifestation of the underlying
vortex condensation.

Finally, the vortices observed in our simulations possess a thick core;
the thickness of the core is the cause for approximate Casimir scaling
of potentials at intermediate distances. 
This solves a long-standing problem of the center vortex theory.

The ``spaghetti vacuum'' picture, believed to be dead for
more than a decade, returns to the stage, in quite a good health.

%
%

\end{document}